\documentclass[aps,prd,10pt,
               preprintnumbers,numbers,sort&compress,nofootinbib,
               twocolumn,
               colorlinks, linkcolor=blue,
               citecolor=blue]{revtex4}
\usepackage{epsfig}

\begin{document}
\title{Atmospheric Radio Signals From Quark Nugget Dark Matter}
\author{Kyle Lawson}
\affiliation{Department of Physics and Astronomy, University of
  British Columbia, Vancouver, BC, V6T 1Z1, Canada}
\email[E-mail:~]{klawson@phas.ubc.ca}

\begin{abstract}
If the dark matter of our galaxy is composed of heavy nuggets of quarks 
or antiquarks in a colour superconducting phase there will be a 
small but non-zero flux of these objects through the Earth's atmosphere. 
A nugget of quark matter will deposit only a small 
fraction of its kinetic energy in the atmosphere and is unlikely to be 
detectable. If however the impacting object is composed of 
antiquarks the energy deposited can be quite large and contain 
a significant charged particle content. These relativistic 
secondary particles will subsequently be deflected by the earth's 
magnetic field resulting in the emission of synchrotron radiation. This 
work argues that this radiation, along with a thermal component emitted 
from the nugget's surface, should be detectable at radio frequencies and that 
both present and proposed experiments are likely to prove capable of  detecting 
such a signal. 
\end{abstract}
\maketitle
\section{Introduction}
\label{sec:intro}
Several recent experiments have  attempted to detect the 
radio emission associated with extensive air showers initiated 
by the impact of an ultrahigh energy cosmic ray on the earth's atmosphere
\cite{Ardouin:2009zp},\cite{Huege:2012vk}, \cite{Kelley:2012as}. 
This radio wavelength emission is generated by the deflection of 
secondary particles by the earth's magnetic field which produces  
synchrotron radiation. It is the purpose of this paper to demonstrate that 
experiments of this type will also capable of placing limits on dark matter 
in the form of heavy  quark matter nuggets. 
In the following section I offer a brief overview of the 
quark nugget dark matter model and it's observational 
consequences for both galactic observations and ground based 
detectors. Section \ref{sec:radio_emis} extends the previous analysis 
of the air shower induced by a quark nugget 
passing through the earth's atmosphere \cite{Lawson:2010uz} and 
estimates the synchrotron signal generated by such an event. 
Sections \ref{sec:Efield_mag} and \ref{sec:intens} then determine specific 
observable properties of the radio emission. Finally, section \ref{sec:detect} 
offers a brief description of fututre detection prospects.
\subsection{dark matter as compact composite objects}
\label{subsec:gal}
The microscopic nature of the dark matter is not yet 
established. While the majority of dark matter models introduce 
a new particle which is fundamentally 
weak in its interactions with visible matter it is only 
the  the dark matter  interaction cross 
section to mass ratio ($\sigma / M$) that is 
observationally constrained. As such, a sufficiently heavy 
dark matter candidate may avoid observational constraints despite 
having a relatively large interaction strength. This is possible due to the 
fact that only the dark matter mass density is directly measured, 
a heavier dark matter candidate will have a lower number 
density and thus smaller flux through any detector. 
In the model to be discussed here the dark matter takes 
the form of heavy compact composite objects with nuclear scale 
densities and composed of the standard light quarks 
\cite{Zhitnitsky:2002qa},\cite{Oaknin:2003uv}. 
Several versions of this model have been proposed dating back 
to objects such as stranglets \cite{Witten:1984rs}.
Current observational constraints
on the dark matter $\sigma / M$ ratio require that these quark nuggets 
carry a mean baryonic charge of at least $10^{23}$. This lower bound 
is taken from a number of direct detection experiments as detailed 
in \cite{Zhitnitsky:2006tu}. There is also a region of parameter space 
at $B\sim 10^{29}  - 10^{31}$ likely excluded based on lunar seismic 
data \cite{Herrin:2005kb}.

Quark nuggets may  be formed at the \textsc{qcd} phase transition 
and remain stable over the lifetime of the universe. Nugget formation 
allows for the creation of nuggets of both matter and 
antimatter and may explain both the dark 
matter and the matter-antimatter asymmetry of the universe.
This is possible if the production of antinuggets is favoured by a 
factor of $\sim 1.5$ over the production of matter nuggets 
leading to dark matter which consists of 
two parts matter nuggets to three parts antimatter nuggets. The 
excess matter (that  not bound in the nuggets) then forms the visible matter of 
the universe in the observed five to one matter to dark matter ratio. 

Originally proposed to explain the baryon asymmetry rather 
than any specific galactic observation this  model 
has been found to have several observational consequences for the 
galactic spectrum. 

${\bullet}$ The quark nuggets are surrounded by an ``electrosphere" 
of positrons the outer layers of which are bound with energies at typical 
atomic scales. The annihilation of these positrons with the electrons 
of the interstellar medium will result in a positronium decay line (and 
associated three photon continuum) in regions where both the visible 
and dark matter densities are large. In particular one should expect 
a $511keV$ line from the galactic centre \cite{Oaknin:2004mn} 
\cite{Zhitnitsky:2006tu}. Such a spectral feature is in fact observed 
and has been studied by the \textsc{Integral} observatory \cite{Jean:2005af}. 

${\bullet}$ Positrons closer to the quark matter surface can carry energies up 
to the nuclear scale. If a galactic electron is able to penetrate to a sufficiently 
large depth it will no longer produce  the characteristic positronium decay 
spectrum but a direct $e^-e^+ \rightarrow 2\gamma$ emission spectrum 
\cite{Lawson:2007kp}. Precisely modeling the transition between these 
two regimes allows for the determination of
the strength of the MeV scale emissions relative to 
that of the 511keV line \cite{Forbes:2009wg}. Observations 
by the \textsc{Comptel} satellite show an excess above the gamma ray 
background predicted from galactic sources at the energy and 
intensity predicted by this model \cite{Strong:2004de}.

${\bullet}$ Galactic protons may also annihilate with the antimatter 
comprising the quark nuggets. The annihilation of a proton within a 
quark nugget will produce hadronic jets which cascade down into 
lighter modes of the quark matter. If the energy in one of these jets reaches the 
quark surface it will excite the most weakly bound positron states near the 
surface. These excited positrons rapidly lose energy to the strong electric field 
near the quark surface. This process results in the emission of Bremsstrahlung 
photons at x-ray energies \cite{Forbes:2006ba}. Observations by the 
\textsc{Chandra} observatory indicate an excess in x-ray emissions from 
the galactic centre in precisely the energy range predicted \cite{Muno:2004bs}.

${\bullet}$ The annihilation of visible matter within the nuggets heats them above the 
background temperature. The thermal spectrum from the nuggets may be 
predicted based on the emission properties of the electrosphere along with the 
annihilation rate at various positions within the galaxy \cite{Forbes:2008uf}.
The majority of this thermal energy is emitted at the eV scale where it 
is very difficult to observe against the galactic background. However the 
emission spectrum will extend down to the microwave scale where it may 
 be responsible for the ``\textsc{wmap} haze" \cite{Finkbeiner:2003im}.  
 
 ${\bullet}$ The same thermal emission that may contribute to the 
 \textsc{wmap} haze associated with the galactic plane must also have 
 been produced in the high density early universe when the isotropic 
 matter density was comparable to that presently seen in  high density 
 regions such as the galactic centre. Taking the 
 nugget temperature scale from the galactic centre and scaling 
 it with the cosmological expansion one predicts an excess in the 
 isotropic radio background beginning a few decades in frequency  
 below the \textsc{cmb} peak \cite{Lawson:2012zu}. The \textsc{Arcade2} 
 experiment show just such a low frequency 
 isotropic radio excess \cite{Fixsen:2009xn}
 which does not currently have an obvious astrophysical source.

The emission mechanisms involved in describing the full nugget 
spectrum span a very large energy range  from the modified thermal 
spectrum in the microwave up to energies associated with nuclear annihilations 
observable as gamma rays.  However, they make no 
significant contribution to the galactic spectrum 
at the GeV scale or above, an energy range across which the FERMI telescope 
has placed significant constraints on a possible dark matter contribution
\cite{Ackermann:2011wa}, \cite{Ackermann:2010rg}, 
\cite{Abdo:2010dk}. 

While the uncertainties in the astrophysical backgrounds involved make an exact 
determination impossible a best fit to the galactic spectrum favours quark 
nuggets with a mean baryon number of roughly $B\sim 10^{25}$. 
\subsection{quark matter in the atmosphere}
While the low number density of quark nuggets required to 
explain the observed dark matter mass density implies a 
flux many orders of magnitude below conventional dark matter 
models they  may still have observational consequences.  
Assuming the nuggets to have roughly 
nuclear scale densities the observational constraints on 
baryon number cited above imply 
a minimum mass of a gram 
and a radius of $\sim 10^{-7}m$. Assuming the local dark matter mass 
density near the galactic plane average ($\rho \sim 1$Gev/cm$^3$) and a mean velocity 
at the galactic scale $v_g \sim 200$km/s one may make an 
order of magnitude estimate the yearly flux of 
nuggets through the atmosphere as,
\begin{equation}
\label{eq:flux}
\frac{dn}{dt ~ dA} = n_N v_g = \frac{\rho_{DM}}{M_N} v_g = 
\left( \frac{B}{10^{25}} \right) km^{-2} yr^{-1}
\end{equation}
this flux is at a level comparable to that of ultra high energy cosmic 
rays near the \textsc{GZK} cuttoff \cite{Abraham:2008ru}. Consequently
the current generation of large scale cosmic ray observatories have a sufficient 
collection area to conduct a meaning search for these objects provided 
their mass is near the lower end of the allowed range. 
Unlike standard cosmic rays the flux of quark nuggets will experience an 
annual variation as the earth moves through the background of galactic matter. 
This annual variation has been used in dark matter detection experiments such 
as \textsc{Dama} \cite{Belli:1999nz}.

In the case of nuggets of quark matter all energy deposited in the 
atmosphere comes through collisional momentum transfer. Relatively 
little energy is deposited and the nugget continues through the atmosphere 
with virtually no change in velocity. The observational prospect for such an 
event are very low. If however the nugget is composed of antiquarks 
(as the majority will be in the model under consideration) atmospheric 
molecules will annihilate on contact with the quark matter surface 
resulting in the release of substantial amounts of energy into the 
surrounding atmosphere. While the total annihilation of a quark nugget with 
$B=10^{25}$ would release $10^{14}$J of energy only a very small fraction of the 
nugget will annihilate in the time it takes to cross the earth's surface. As the nuggets 
carry sufficient momentum to traverse the entire atmosphere the limiting 
factor is the amount of atmospheric mass which they encounter. 

This energy will be deposited in the form of 
thermal radiation from the nugget as well as relativistic particles and
high energy gamma rays
emitted in the nuclear interactions. The prospects for direct detection of 
secondary particles was discussed in \cite{Lawson:2010uz}. This work 
focusses instead on the prospect of radio frequency detection. 
Both the thermal spectrum and the 
geo-synchrotron emission from the emitted particles will contribute 
to the radio spectrum, the magnitude and relative strength of these 
two emission mechanisms  will be approximated 
in section \ref{sec:radio_emis}. 

The total scale of the radio band spectrum is determined by the 
rate at which atmospheric molecules are annihilated within the 
nugget. This rate was estimated in \cite{Lawson:2010uz} where 
it was shown to increase exponentially with the growing 
atmospheric density until the point at which the thermal energy produced 
by annihilations is sufficient to deflect further incoming matter. At this point the 
annihilation rate reaches an equilibrium point and does not increase 
further. The exact value of the equilibrium rate depends on details 
of emission from the quark surface but should occur when the 
surface temperature is on the order of $10keV$. For this temperature 
scale the annihilation rate saturates a few kilometers above the earth's 
surface. Above this height the annihilation rate is simply determined 
by atmospheric density and the physical cross section of the nugget. 
\begin{eqnarray}
\label{eq:an_rate}
\Gamma_{an}&=& \sigma_{N} v_{N} n_{at}(h) ~~~ h > h_{eq} \\
 &=&  \sigma_{N} v_{N} n_{at}(h_{eq}) ~~~ h < h_{eq} \nonumber  
 \end{eqnarray} 
 Where $\sigma_N$ is the physical cross section of the nugget (on the 
 order of $10^{-10} cm^2$ for the nugget mass range considered here), 
 and $h_{eq}$ is the height at which the equilibrium annihilation rate 
 is reached.
\section{Radio frequency emission}
\label{sec:radio_emis}
Emission in the radio band arrises through two distinct processes: 
thermal emission from the surface of the quark nugget as it is heated 
by annihilations and geo-synchrotron emission from relativistic 
charged particles generated in nuclear annihilations. 
\subsection{thermal emission}
A quark nugget passing through 
the atmosphere will generate a thermal spectrum  
extending across a wide range of energies including the radio band. 
The thermal emission spectrum was first calculated for 
nuggets in the galactic centre \cite{Forbes:2008uf}
where the thermal energy emitted at frequency $\omega$ by 
a nugget of temperature $T$ was calculated to be,
\begin{eqnarray}
\label{eq:therm_spec}
\frac{dE}{dt~d\omega~dA} = &\frac{3}{45\pi}& \alpha^{5/2} T^3
\sqrt[4]{\frac{T}{m_e}}\left( 1 + \frac{\omega}{T}  \right) \nonumber \\ 
&\times& e^{-\omega/T} \left(17 - 12ln(\frac{\omega}{2T}) \right)
\end{eqnarray}
This spectrum is modified from a simple black body spectrum by the 
presence of an electrosphere surrounding the quark matter and is 
suppressed with respect to black body radiation for temperatures below the 
electron mass ($m_e$).  The nugget temperature is determined by the rate at 
which matter is annihilated within the nugget and, as such, will be much higher 
for a nugget in the earth's atmosphere than for nuggets in the interstellar medium. 
The evolution of a nugget's temperature as it moves through the atmosphere 
was described in \cite{Lawson:2010uz} where it was found that the temperature 
typically peaks in the 10s of keV range. At this temperature the emission of radiation in the 
radio band is well described by the $\omega < T$ limit and the spectrum 
may be taken as 
\begin{eqnarray}
\label{eq:therm_spec_large_T}
\frac{dE}{dt~d\omega} &\approx& \left(10^{-10} J s^{-1} Mhz^{-1} \right) \\ 
&\times &\left( \frac{T}{10 keV} \right)^{13/4} \left( \frac{R_n}{10^{-5}cm} \right)^2
~ ln(\frac{2T}{\omega} ) \nonumber
\end{eqnarray}
In this form it is clear that thermal emission generates a relatively flat 
(log dependance) spectral contribution across all radio frequencies. 
Thermal emission will occur uniformly over the nugget surface so that the 
intensity of the thermal component of the spectrum is simply obtained by 
dividing expression (\ref{eq:therm_spec_large_T}) by $4\pi$ 
multiplied by  the square of the 
distance between the nugget and the observer. 
\subsection{geo-synchrotron emission}
\label{sec:geo_sync}
As a quark nugget moves through the atmosphere nuclear annihilations 
generate a large number of secondary particles. Some fraction of these
particles, dominated by relativistic muons, escape from the nugget into  the atmosphere. 
These secondary particles are deflected by earth's magnetic field generating 
synchrotron radiation. 
Note that, as acceleration is perpendicular to velocity, 
the particle velocity remains constant so long as energy losses 
through synchrotron emission and interactions with the atmosphere 
remain small \footnote{As a consistency check it may be noted that 
the total emitted synchrotron radiation, given by integrating expression 
\ref{eq:poynting} over all angles and the lifetime of the muon, is well 
below the initial kinetic energy}.
For relativistic particles in the earth's magnetic field the synchrotron radius 
is much longer than the total path length so that the the deflection of the 
particles is relatively small.  In this limit we can linearize the equations of motion of 
a particle within a constant magnetic field.
\begin{equation}
\label{eq:charge_eom}
\vec{v}(t) \approx \vec{v}_0 +  \dot{\vec{v}}_0 t 
\approx \vec{v}_0 + \left( \frac{q}{m\gamma} \right) \vec{v}_0 \times \vec{B}_0
\end{equation}
Where $v_0$  is the initial velocity of the charged particle which, in 
the case of muons produced in QCD scale processes at the quark 
surface should be on the order of $v_0 \sim 0.9c$. The acceleration term 
in this expression will lead to the production of synchrotron radiation. 
The resulting radiation is beamed along the direction of the velocity 
so that (in the small deflection limit) we need only consider the radiation emitted 
by particles with initial velocities directed towards the observer. 

Unlike a conventional air shower the dominant mechanism generating 
relativistic particles is not direct pair production but complex many body 
interactions at the quark matter surface. In the model 
considered here these processes may be summarized by two simple parameters
the mean velocity of the muons emitted ($\beta_0$) and the number of muons 
emitted per nuclear annihilation ($f_{\mu}$). 
While they could in principle be calculated within a particular quark matter 
model the introduction of these parameters, intended to capture only
the approximate scale of the radio signal, allows for a discussion of the  
general properties of the synchrotron emission without detailed 
and model specific nuclear calculations. 

It is the uncertainty in these two parameters which will dominate 
the uncertainties in all the calculations which follow. The total number of 
muons directly sets the overall scale of the resulting radio signal produced 
by geo-synchrotron effects. As such this scale is estimated here only 
at the order of magnitude level. While the general shape of the resulting 
radio spectrum is relatively robust its exact details are dependent on the 
energy distribution of the muons escaping from the nugget. Both the 
total emission rate and the energy spectrum depend on the efficiency 
of energy transfer through the quark matter of the nugget. The calculation 
of the mode and efficiency of energy transfer in high density QCD is a 
complicated and phase dependent problem beyond the scope of this paper 
which seeks only to demonstrate the feasibility of detecting quark nugget induced 
air showers rather than to make highly specific calculations of the resultant 
emission spectrum. 

Muons produced through 
nuclear interactions within the quark nugget necessarily carry nuclear 
scale energies and thus $\beta_0 \sim 0.9 - 0.99$. The muon production 
coefficient is more difficult to estimate but, in the simplified picture where 
a nuclear annihilation in the quark matter produces a muon with properties 
relatively close to their vacuum values, it may be estimated that only those 
muons emitted in the direction of the surface will escape (the 
remainder being thermalized within the nugget.) The sharpness of the 
quark surface will also result in a high likelihood of an outgoing 
muon being reflected back into the quark matter. The combination of these 
two effects suggests a muon production coefficient on the order 
$f_{\mu} \approx 0.1 - 0.01$. 

While muon production does not occur through direct pair production 
the process is charge independent so $\mu^+$ and $\mu^-$ production 
proceed at the same rate. In this case it is relatively simple to evaluate the 
electric field generated by a charge neutral muon pair in the small $\omega_B$ 
limit. This field at leading order is, 
\begin{equation}
\label{eq:Efield}
|\vec{E}(\vec{r},t)| = \frac{q}{2\pi\epsilon_0} \frac{\omega_B}{c R(t)}
\frac{ \beta_0 sin\theta_{vB}}{\gamma (1-\beta_0)^2} 
\end{equation}
Where $sin\theta_{vB}$ is the angle between the initial muon velocity 
and the earth's magnetic field and $R(t)$ is the distance between the muon 
and the observation point. 
This may be transformed into momentum space as,
\begin{equation}
\label{eq:Efield_w}
|\vec{\mathcal{E}}(\vec{r},\omega)| = \frac{q}{(2\pi)^{3/2} \epsilon_0} 
\frac{\omega_B sin\theta_{vB}}{c^2 \gamma (1-\beta_0)^2} |\mathcal{I}(R,\omega))| 
\end{equation}
Where I have defined the integral
\begin{eqnarray}
\label{eq:angle_int}
|\mathcal{I}(R,\omega)|^2 &=& \left( \int_{\Delta R/R_0}^1 \frac{dx}{x} 
cos\left( \frac{R_0 \omega}{c \beta_0} x \right) \right)^2  \nonumber \\
&+& \left( \int_{\Delta R/R_0}^1 \frac{dx}{x} 
sin\left( \frac{R_0 \omega}{c \beta_0} x \right) \right)^2 
\end{eqnarray}
With $R_0$ being the nugget to observer distance and $\Delta R$ the smallest 
separation between the emitting muon and the observer. Note that the 
relevant timescale in determining the frequency dependence of the 
synchrotron radiation is the length of time for which the emitting particle
pair remains relativistic. It should also be noted that all information about the 
shower geometry is carried by the sine function and the unitless integral. 
This expression then allows us to estimate the scale of the field strength based 
purely on the numerical coefficient. 
\begin{eqnarray}
|\vec{\mathcal{E}}(\vec{r},\omega)| \approx
&(&5\times 10^{-10} \mu V ~m^{-1}~Mhz^{-1}) \nonumber \\
&\times&  \left( \frac{B}{0.5G} \right) \frac{sin\theta_{vB} |\mathcal{I}|}{\gamma (1-\beta)^2}   
\end{eqnarray} 
As should be expected the contribution of a single particle pair 
is relatively weak. In order to determine the total field strength 
associated with the shower this value must be scaled up by the 
total number of particles contributing to the radio emission
along a given line of sight. 

The basic properties of a nugget induced air shower can be 
demonstrated in the geometrically simplified case where
the quark nugget moves through the atmosphere vertically. In this 
case the earth's magnetic field may be taken to lie in the x-z plane with 
an inclination angle $\theta_B$ and the observation position on the 
earth's surface by a distance from the shower centre ($b$) and a angle 
relative to the horizontal component of the magnetic field ($\phi$). 
Due to the beaming of the synchrotron radiation only muons 
with an initial velocity directed very nearly towards the observer need 
be considered so the particle to observer distance may be approximated as 
$R(t) \approx R_0 - \beta_0 c t$ with $R_0 = \sqrt{h^2+b^2}$ where $h$ 
is the nugget's height at the time the muon is emitted. This 
approximation breaks down if the muons actual reach the detector, and 
a short distance cutoff must be added for points 
sufficiently close to the shower core.  In the following calculations 
the integral \ref{eq:angle_int} will be cutoff when the minimum 
separation distance ($\Delta R$) becomes comparable to the 
separation distance of the muon pair.
\section{Electric field Magnitude} 
\label{sec:Efield_mag}
\begin{figure}[t]
\begin{center}
\includegraphics[width = 0.4\textwidth]{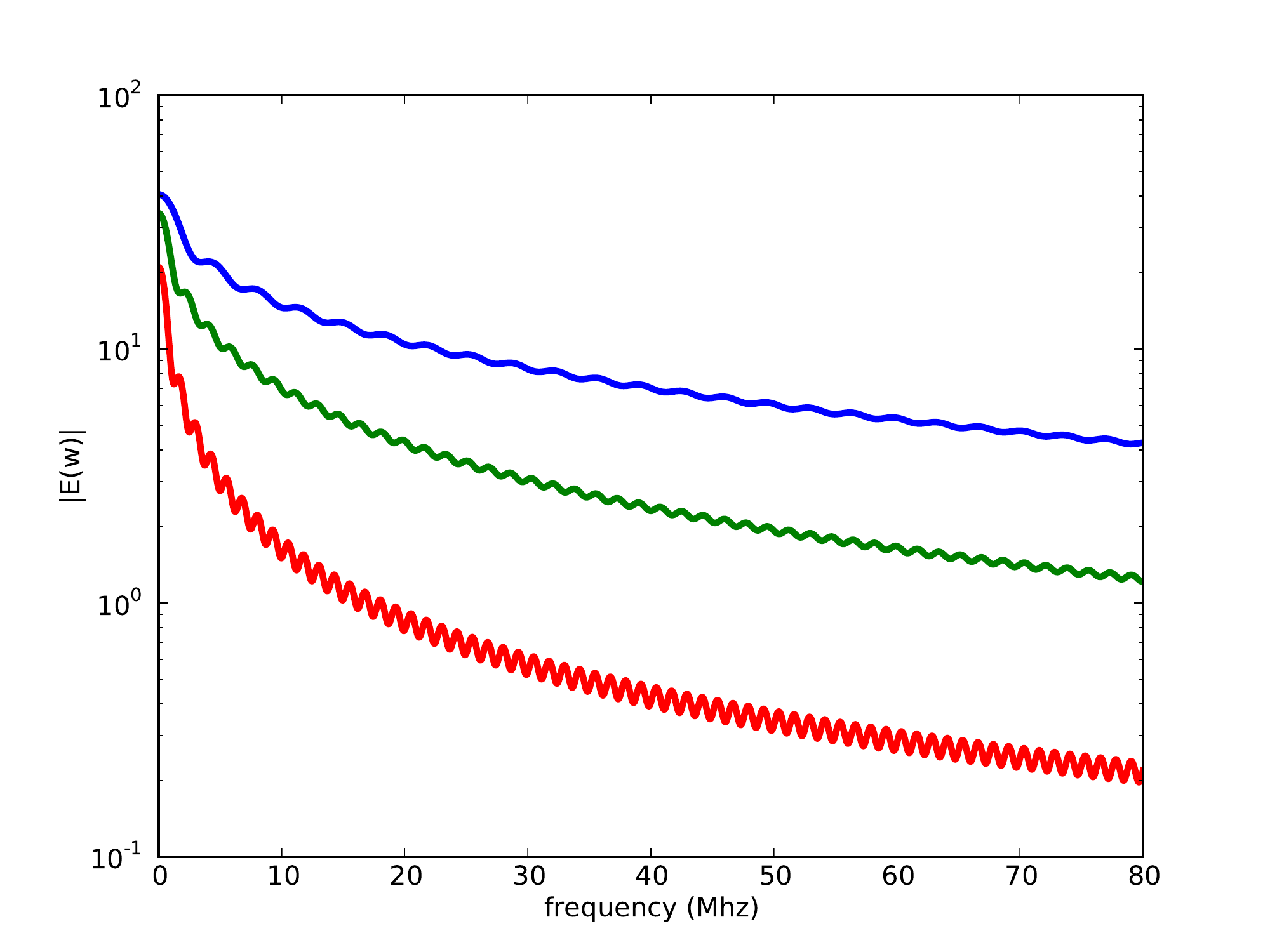}
\caption{Total electric field strength generated by geosynchrotron emission 
in units of $\mu V~ m^{-1} Mhz^{-1}$
as observed 250m from the shower core when the 
nugget is at height h=500m (blue), h = 1km (green), h = 1500m (red).}
\label{fig:Efield}
\end{center} 
\end{figure}

\begin{figure}[t]
\begin{center}
\includegraphics[width = 0.4\textwidth]{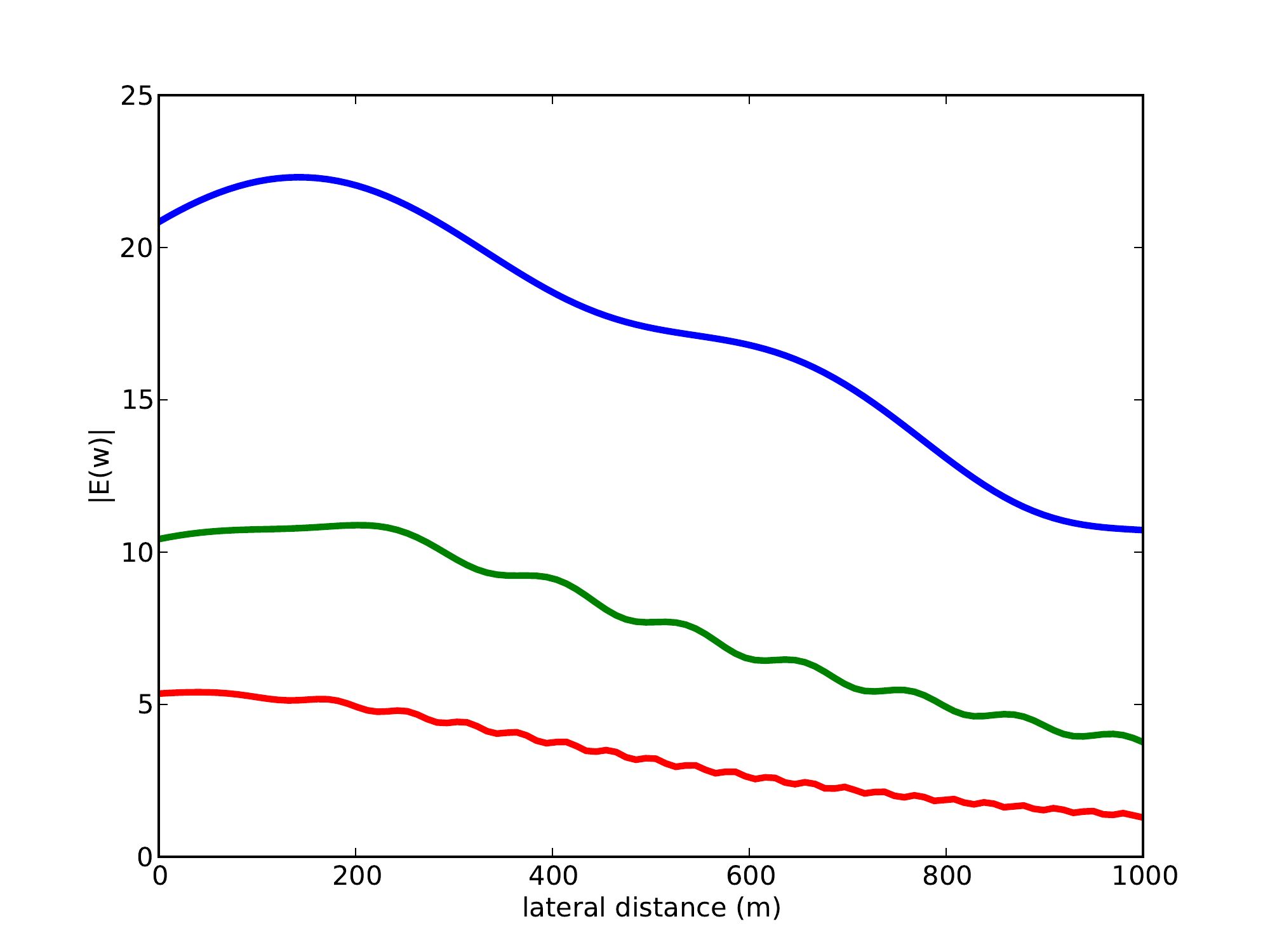}
\caption{Total electric field strength generated by geosynchrotron emission 
in units of $\mu V~ m^{-1} Mhz^{-1}$
at different lateral distances from the shower core. The different 
contours are at frequencies $\omega = 5Mhz$ (blue), 
$\omega = 20Mhz$ (green), $\omega = 60Mhz$ (red).}
\label{fig:lat}
\end{center} 
\end{figure}

The expression (\ref{eq:Efield_w}) gives the momentum space electric 
field strength resulting from a single muon pair. In order to determine the 
total field strength at the earth's surface we now sum over all muon pairs 
emitted towards the observer at a given moment. The nuclear annihilation 
rate for a given height is given in expression (\ref{eq:an_rate}) so that 
the total number of muons produced is determined by multiplying this rate by  the 
muon production coefficient ($f_{\mu}$) as discussed above. The rate of 
muon production from a specific point on the nugget surface depends on 
the local flux of material onto the quark surface which is proportional to 
the area perpendicular to the nugget's velocity.
\begin{equation}
\label{eq:dN_dOmega}
\frac{dN_{\mu}}{d\Omega ~ dt} = \frac{\Gamma_{an}}{2\pi} f_{\mu} cos\phi 
\end{equation}
Here $\phi$ is the angle between the direction of emission and the nugget's
velocity. As stated above, muon emission is dominantly perpendicular to the 
quark surface so that the angular position of emission on the nugget's surface 
fully determines the direction of the resulting synchrotron radiation. As such, 
in the vertical shower case, the radiation detected a distance $b$ from the 
shower centre when the nugget is a a height $h_N$ has  
\begin{equation}
\label{eq:dN_dOmega_b}
\frac{dN_{\mu}}{d\Omega ~ dt} = \frac{\Gamma_{an}}{2\pi} 
\frac{h_N}{\sqrt{h_N^2 + b^2}} 
\end{equation}
In order to determine the total number of muons contributing to the 
observed flux we must estimate the solid angle at the nugget's surface 
which contributes to the synchrotron radiation along a given line 
of sight. To do this we note that the intensity of the radiation has an angular 
dependence which scales as $S \sim (1 - \beta cos\theta )^4$ which, 
for $\beta$ close to one, is sharply peaked around zero. As such I  
take as the angular scale of emission the angle at which the intensity 
falls to half its peak value. This angle is defined by the expression
\begin{eqnarray}
\label{eq:half_angle}
\left( \frac{1-\beta}{1-\beta cos\theta_{1/2}}\right)^4 = \frac{1}{2} \nonumber \\
\theta_{1/2}^2 \approx 0.38 \left( \frac{1-\beta}{\beta} \right)
\end{eqnarray}
Where the second expression uses the small angle approximation.
The relevant solid angle of the nugget surface is then given by
$d\Omega \approx \theta_{1/2}^2$ and the total number of muons 
emitted towards the observer at a given time is,
\begin{equation}
\label{eq:obs_muons}
\frac{dN_{\mu}}{dt} = \Gamma_{an} f_{\mu} \frac{\theta_{1/2}^2}{2\pi} 
\frac{h_N}{\sqrt{h_N^2 + b^2}} 
\end{equation}
Finally we note that the nugget's velocity is much smaller than 
that of the emitted muons so that its height changes very little over 
the time scale on which the muons emit synchrotron radiation. This 
time scale depends on interaction rates with the surrounding atmosphere 
and a full calculation of energy loss rate for a muon is rather complicated. 
In order to estimate the time scale involved I will simply note that muons 
lose energy to the surrounding atmosphere much slower than electrons 
and assume that the muon energy is roughly constant until it decays to an 
electron at which point it is rapidly stopped as it scatters off atmospheric 
molecules. In this approximation the muon lifetime sets the timescale over which the 
emitting particle remains relativistic \footnote{This is a serious simplification 
of the very complicated propagation of the secondary particles of an extensive 
air shower a full treatment of which would involve far more detailed simulations. 
However, 
the uncertainty inherent in the initial particle production rate and energy spectrum is 
great enough that a detailed numerical treatment is not warranted at present.}. 
Under these approximation the number of 
muons contributing to the radiation flux when the nugget is at a given height
may be approximated as, 
\begin{equation}
\label{eq:total_obs_muons}
N_{\mu}(h_N) = \Gamma_{an} f_{\mu} \gamma \tau_{\mu} 
\frac{\theta_{1/2}^2}{2\pi} \frac{h_N}{\sqrt{h_N^2 + b^2}} 
\end{equation}
where $\tau_{\mu}$ is the muon lifetime and $\gamma$ is 
the initial muon boost factor. Multiplying this expression by 
the electric field contribution from a single muon pair 
(\ref{eq:Efield_w}) will then give the total field strength 
when the nugget is at a given height. Figure \ref{fig:Efield} 
shows the electric field strength generated by geosynchrotron 
emission when the nugget is at different heights. Figure 
\ref{fig:lat}  shows the lateral profile of electric field strength as 
received at the surface. Note that the oscillations appearing 
in the field strength arise from the unphysical assumption that all muons 
are emitted at the same energy. Averaging over a range of 
initial muon energies would erase this effect but would not change  
the scale of the emission or the basic form of the spectrum. 

\section{Total intensity}
\label{sec:intens}
\begin{figure}[t]
\begin{center}
\includegraphics[width = 0.4\textwidth]{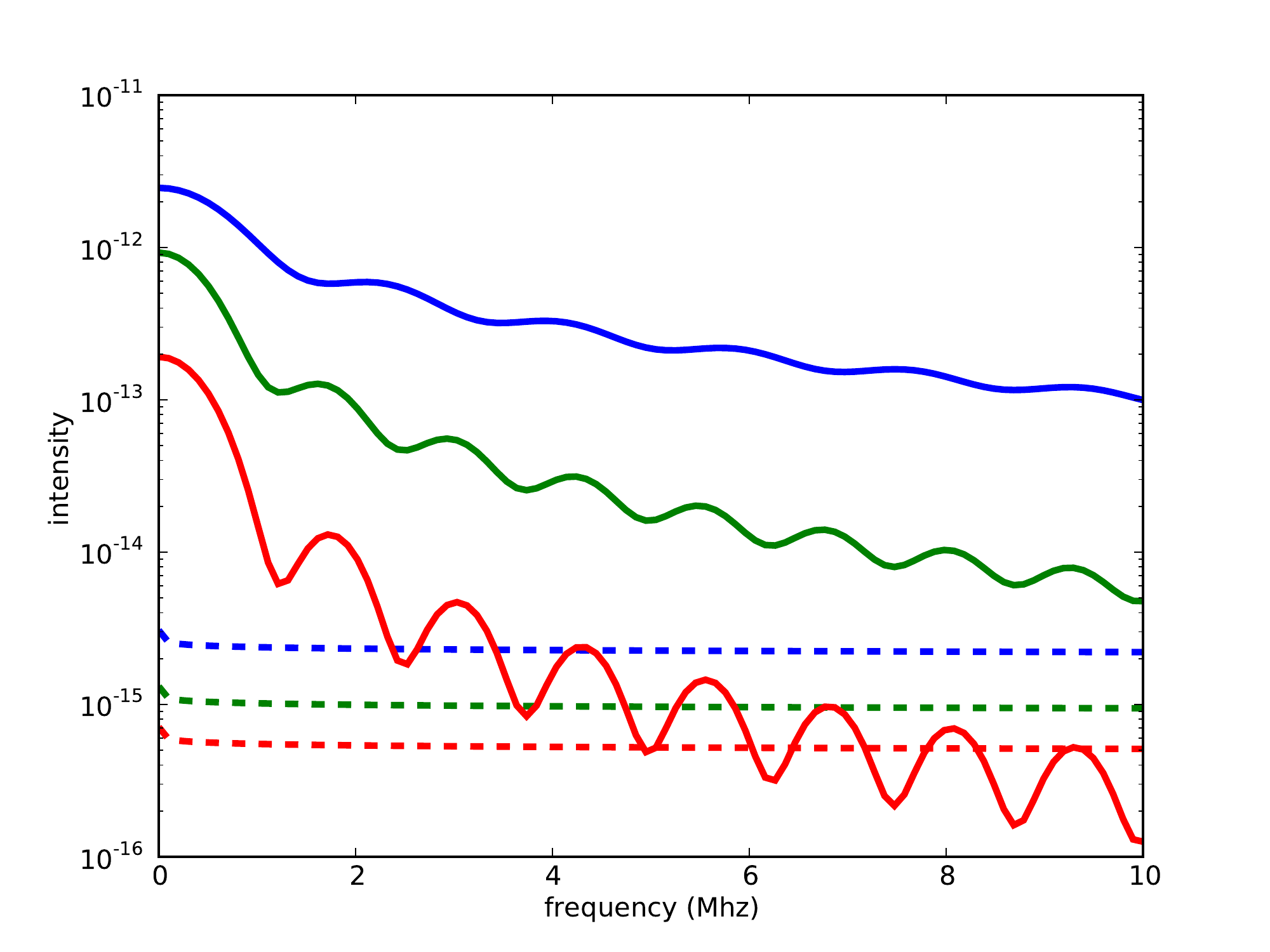}
\caption{Total intensity ($J s^{-1} m^{-2} Mhz^{-1}$) 
received on the shower axis (b=0) 
when the nugget is at heights 1km, 1.5km and 2km (blue, green 
red respectively.) In each case the thermal intensity (dashed) 
is found to be below the 
contribution from geo-synchrotron radiation (solid).}
\label{fig:tot_in}
\end{center} 
\end{figure}

Finally I determine the total intensity resulting from the 
passage of a nugget through the atmosphere. The magnitude of 
the Poynting vector is given by,
\begin{equation}
\label{eq:poynting}
|\vec{S}| = \frac{dE}{dt~dA} = \frac{\vec{E}^2}{\mu_0 c}
\end{equation}
Or, in momentum space
\begin{equation}
\label{eq:poynting_w}
\frac{dE}{d\omega~dA} = \frac{|\mathcal{E}(\omega)|^2}{\mu_0 c}
\end{equation}
With the  field strength for a single muon pair given by (\ref{eq:Efield_w}).
This radiation is emitted as long as the muon remains relativistic, as argued 
above this timescale may be roughly estimated by the observer frame 
muon lifetime. In this case the observed flux per charge pair can be 
approximated as, 
\begin{equation}
\label{eq:single_mu_flux}
\frac{dE}{d\omega~dt~dA} = \frac{1}{\gamma \tau_{\mu}}\frac{|\mathcal{E}|^2}{\mu_0 c}
\end{equation} 
The electric field generated by all particles moving towards the observer 
at a given time add coherently so that this may be translated to a total 
flux from all particles by multiplying the field strength by the total number 
of contributing muon pairs which is half the value determined in 
(\ref{eq:total_obs_muons}).
\begin{eqnarray}
\label{eq:tot_synch_flux}
\frac{dE}{d\omega~dt~dA} &=& \frac{1}{\gamma \tau_{\mu}} 
\frac{|N_{\mu}\mathcal{E}|^2}{\mu_0 c}  \\
&=& \left(10^{-16} J m^{-2} s^{-1} Mhz^{-1} \right) 
\left(\frac{\Gamma_{an} f_{\mu}}{10^{16}s^{-1}}\right)^2 \nonumber \\
&\times& \left( \frac{B}{0.5G}\right)^2 \frac{\gamma ~ |\mathcal{I}|^2}{\beta^2(1-\beta)^2} 
~\left(\frac{h_N^2}{b^2+h_N^2}\right) \nonumber 
sin^2\theta_{vB}
\end{eqnarray}
I want to add to this the intensity contribution from the thermal spectrum 
(\ref{eq:therm_spec}). As the thermal radiation is emitted uniformly from the 
nugget's surface the total intensity at a given position is simply given by 
\begin{equation}
\label{eq:therm_intens}
\frac{dE}{d\omega ~dt ~dA} = \frac{1}{4\pi (b^2 + h_N^2)} \frac{dE}{d\omega~dt}
\end{equation}
The intensity of both the thermal and synchrotron radiation components 
are plotted in
\ref{fig:tot_in}. In the range of frequencies relevant to most current and planned
experiments the thermal emission has only a minimal impact on the total 
intensity. If however observations are extended into the Ghz range the thermal 
component, with its relatively flat spectrum will become increasingly important. 

\begin{figure}[t]
\begin{center}
\includegraphics[width = 0.4\textwidth]{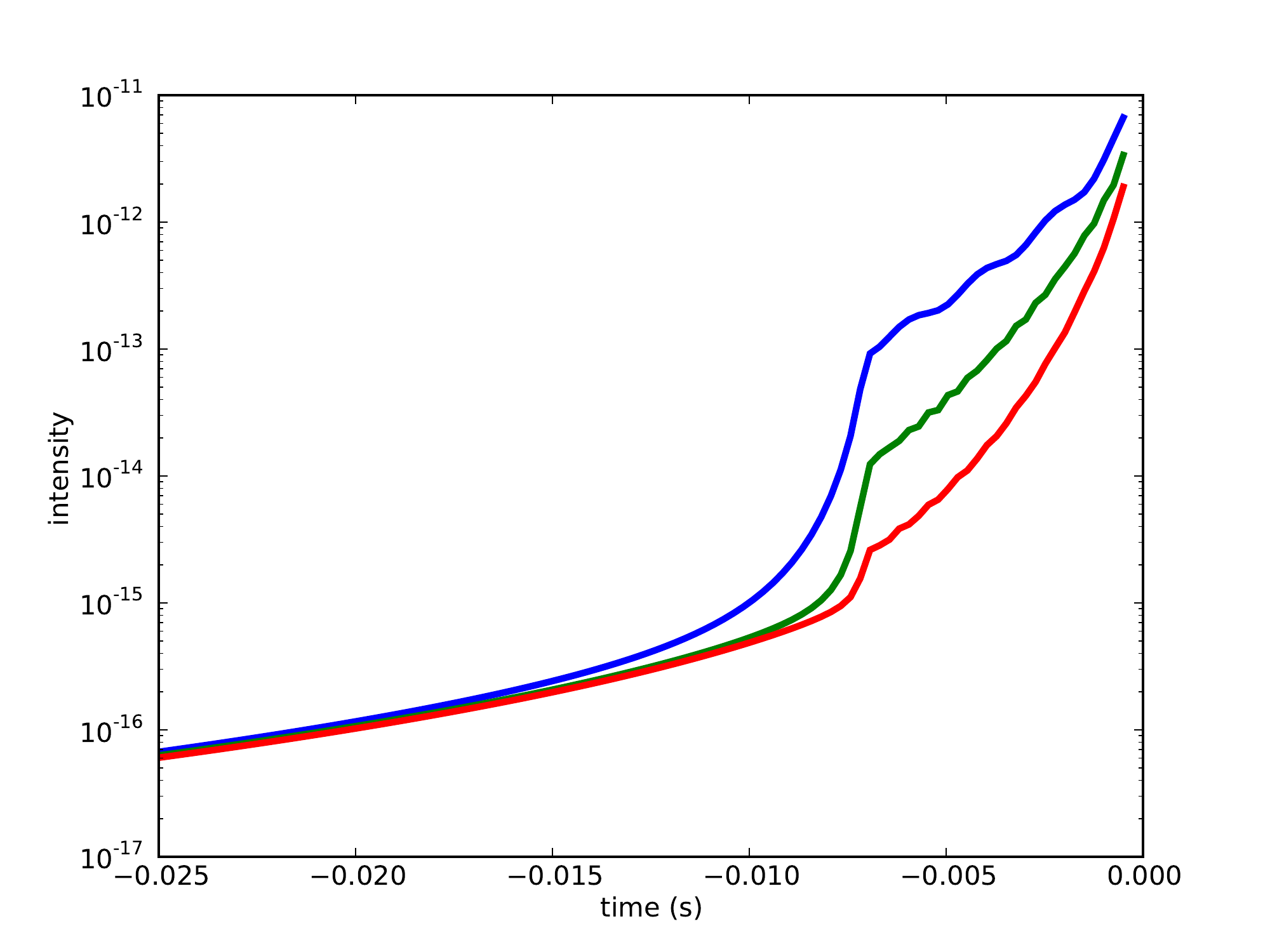}
\caption{Total intensity ($J s^{-1} m^{-2} Mhz^{-1}$) from 
both thermal and geosynchrotron 
received on the shower axis (b=0) as a function of time
where the nugget is taken to reach the surface at t=0. 
Intensity profiles are shown at 5Mhz (blue), 20Mhz(green) and
60Mhz (red).}
\label{fig:time_dep}
\end{center} 
\end{figure}

Finally, it should be noted that the delivered intensity evolves on a 
timescale much longer than that generated by an air shower triggered 
by a single ultrahigh energy proton or nuclei. The timescale in this model 
is set by the time it takes for the quark nugget to pass through the earth's 
atmosphere. Assuming that the quark nuggets carry velocities at a typical 
galactic scale of $\sim 200km/s$ the rise in intensity will take place over 
tens of milliseconds. Figure \ref{fig:time_dep} shows the rise in intensity 
as a function of time. 

\section{Detection Prospects} \label{sec:detect}
Having established the basic properties of the radio band emission 
generated by an antiquark nugget passing through the atmosphere 
it is possible to speculate on the possibility for such an event to be 
observed using present or planned detectors. 
In this context two features of the emission are of particular 
relevance: the total emission strength and the lateral extent of the 
emission at the surface. 

As shown in the lateral profile plotted in figure \ref{fig:lat} 
the radio band emission extends out to kilometer scales. As such 
the these events may be observed over a extended area surrounding 
the shower core. Of particular interest in this context are radio 
detection facilities associated with large scale cosmic ray detectors. 
The coincident arrival a radio pulse with a millisecond duration 
and a particle shower would be a strong smoking gun 
signal as all known phenomenon capable of generating a large number 
of secondary particles evolve on much shorter (nanosecond) timescales. 

Of particular interest here are experiments such as the Auger Engineering 
Radio Array (\textsc{AERA}), \textsc{LOPES} 
at the \textsc{KASCADE}-Grande array, and \textsc{CODALEMA}. 
Each of these radio detection experiments have a sufficient spacial 
extent to observe kilometer scale events and a sensitivity to the 
tens of Mhz range across which both the geosynchrotron and 
thermal emission  generated by the nugget will extend. See 
\cite{Ardouin:2009zp}, \cite{Huege:2012vk}, \cite{Kelley:2012as} 
and references therein for details of these experiments. 

Provided that timing cuts do not remove this type of long duration 
radio pulse these experiments should be very capable of setting 
strict limits on the flux of antiquark nuggets in the mass range 
favoured by fits to the galactic emission as discussed in 
section (\ref{subsec:gal}) as they accumulate further data.

It has also been suggested that the balloon-borne \textsc{ANITA} 
experiment \cite{Gorham:2008dv}  may be sensitive to the thermal 
emission generated by the passage of an antiquark nugget through the 
the radio transparent Antarctic ice \cite{Gorham:2012hy}. Data currently 
under analysis by the \textsc{ANITA} collaboration is likely to place 
constraints on the quark nugget flux across much of the  
preferred mass range. 

\section{conclusion}
This paper has aimed to demonstrate the feasibility of detecting the 
presence of dark matter in the form of heavy nuggets of quark matter
using radio frequency detectors. The passage of a quark nugget 
through the atmosphere will induce an extensive air showers involving 
many secondary charged particles which subsequently emit 
synchrotron radiation across the Mhz band
as they are deflected by the earth's magnetic field. The resulting 
radio emission is likely to be detectable up to a few kilometers from the 
shower core. As such these events should be readily detectable by 
experiments intended to observe radio emission from ultra high energy 
cosmic rays. 

While the intensity and event rate of these showers may 
be at a similar scale to that of air showers initiated by a 
single ultra high energy proton or nucleus 
antiquark nugget induced events have several easily observable distinguishing 
properties.

The nuggets require a time scale on the order of tens of milliseconds 
to traverse the earth's atmosphere and will generate a synchrotron 
signal over much of this time. As such the radio signal will be have a 
duration much longer than that of typical cosmic ray events which evolve 
on time scales orders of magnitude faster. 

The nuggets carry galactic scale velocities and
their flux will show a seasonal variation, however, any microscopic particle 
with so small a velocity will be insufficiently energetic to initiate an 
extensive air shower. Consequently, the detection of a seasonal variation 
in the air shower rate would be a strong indicator of a quark nugget 
contribution to this flux.  

The composite nature of the primary particle in a quark nugget 
initiated air shower means that there is a thermal component to the 
spectrum in addition to the geosynchrotron emission generated by the 
secondary particles. While the thermal component is significantly lower 
than the synchrotron signal at the frequencies typically observed it's 
contribution increases at higher frequencies and may well be observable 
as distinct from the synchrotron signal. 

If, as argued above, large scale cosmic ray detectors are also capable of 
observing the air showers induced by dark matter in the form of 
heavy nuggets of quark matter than these properties will allow the 
two components to be readily distinguished through observations 
at radio frequencies.

\section{Acknowledgments}
I would like to thank Ariel Zhitnitsky for many helpful discussions. 
I would also like to thank Peter Gorham for calling my attention to the 
relevance of the \textsc{ANITA} experiment in this context. 
This research was supported in part by the Natural Sciences and Engineering 
Research Council of Canada and the UBC Doctoral Fellowship program. 

\bibliographystyle{h-physrev}
\bibliography{cosmicrays}

\end{document}